\documentclass{SciPost}

\hypersetup{
    colorlinks,
    linkcolor={red!50!black},
    citecolor={blue!50!black},
    urlcolor={blue!80!black}
}

\usepackage[bitstream-charter]{mathdesign}
\DeclareSymbolFont{usualmathcal}{OMS}{cmsy}{m}{n}
\DeclareSymbolFontAlphabet{\mathcal}{usualmathcal}

\fancypagestyle{SPstyle}{
\fancyhf{}
\lhead{\colorbox{scipostdeepblue}{\bf \color{white} ~SciPost Physics Proceedings }}
\rhead{{\bf \color{scipostdeepblue} ~Submission }}

\fancyfoot[C]{\textbf{\thepage}}
}

\usepackage{hyperref}       
\usepackage{url}            
\usepackage{booktabs}       
\usepackage{cleveref}       
\usepackage{subcaption}     
\usepackage[export]{adjustbox} 
\usepackage[table,xcdraw]{xcolor}
\usepackage{orcidlink}      

\begin{document}

\pagestyle{SPstyle}

\begin{center}{\Large \textbf{\color{scipostdeepblue}{
Physics Informed Neural Networks for design optimisation of diamond particle detectors for charged particle fast-tracking at high luminosity hadron colliders \\
}}}\end{center}

\begin{center}\textbf{
Alessandro Bombini\orcidlink{0000-0001-7225-3355}\textsuperscript{1,2$\dagger$}, 
Alessandro Rosa\orcidlink{0009-0001-3253-5013}\textsuperscript{1}, 
Clarissa Buti\orcidlink{0009-0009-2488-5548}\textsuperscript{1},
Giovanni Passaleva\orcidlink{0000-0002-8077-8378}\textsuperscript{1} and
Lucio Anderlini\orcidlink{0000-0001-6808-2418}\textsuperscript{1} 
}\end{center}

\begin{center}
{\bf 1} Istituto Nazionale di Fisica Nucleare, Sezione di Firenze, Via B. Rossi 1, 50019 Sesto Fiorentino (FI), Italy
\\
{\bf 2} ICSC - Centro Nazionale di Ricerca in High Performance Computing, Big Data \& Quantum Computing, Via Magnanelli 2, 40033, Casalecchio di Reno (BO), Italy
\\[\baselineskip]
$\dagger$ \href{mailto:bombini@fi.infn.it}{\small bombini @  fi.infn.it}
\end{center}

\definecolor{palegray}{gray}{0.95}
\begin{center}
\colorbox{palegray}{
  \begin{tabular}{rr}
  \begin{minipage}{0.37\textwidth}
    \includegraphics[width=60mm]{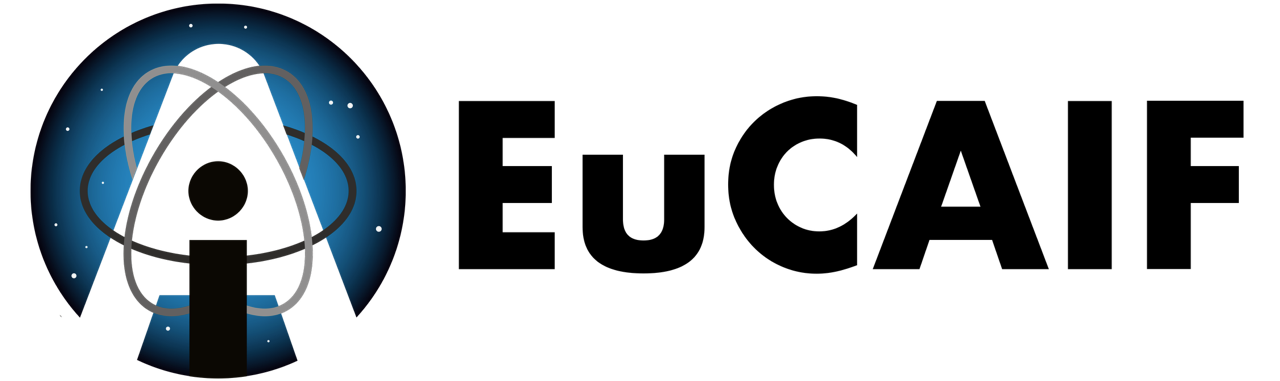}
  \end{minipage}
  &
  \begin{minipage}{0.5\textwidth}
    \vspace{5pt}
    \vspace{0.5\baselineskip} 
    \begin{center} \hspace{5pt}
    {\it The 2nd European AI for Fundamental \\Physics Conference (EuCAIFCon2025)} \\
    {\it Cagliari, Sardinia, 16-20 June 2025
    }
    \vspace{0.5\baselineskip} 
    \vspace{5pt}
    \end{center}
    
  \end{minipage}
\end{tabular}
}
\end{center}

\section*{\color{scipostdeepblue}{Abstract}}
\textbf{\boldmath{%
Future high‑luminosity hadron colliders demand tracking detectors with extreme radiation tolerance, high spatial precision, and sub‑nanosecond timing. 3D diamond pixel sensors offer these capabilities due to diamond’s radiation hardness and high carrier mobility. Conductive electrodes, produced via femtosecond IR laser pulses, exhibit high resistivity that delays  signal propagation. This effect necessitates extending the classical Ramo–Shockley weighting potential formalism. We model the phenomenon through a 3rd‑order, 3+1D PDE derived as a quasi‑stationary approximation of Maxwell’s equations. The PDE is solved numerically and coupled with charge transport simulations for realistic 3D sensor geometries. A Mixture‑of‑Experts Physics‑Informed Neural Network, trained on Spectral Method data, provides a meshless solver to assess timing degradation from electrode resistance. 
}}

\vspace{\baselineskip}

\noindent\textcolor{white!90!black}{%
\fbox{\parbox{0.975\linewidth}{%
\textcolor{white!40!black}{\begin{tabular}{lr}%
  \begin{minipage}{0.6\textwidth}%
    {\small Copyright attribution to authors. \newline
    This work is a submission to SciPost Phys. Proc. \newline
    License information to appear upon publication. \newline
    Publication information to appear upon publication.}
  \end{minipage} & \begin{minipage}{0.4\textwidth}
    {\small Received Date \newline Accepted Date \newline Published Date}%
  \end{minipage}
\end{tabular}}
}}
}



\newpage
\section{Introduction} \label{sec:intro}



Designing detectors for next-generation particle physics experiments demands reliable performance under extreme particle fluxes, driven by the pursuit of high instantaneous luminosity. This challenge is central to upgrades for High-Luminosity LHC~\cite{DaVia:2012ay, CMSTracker:2019aux, DallaBetta:2016czx, DallaBetta:2016dqn}, LHCb~\cite{LHCB-TDR-023, LHCB-TDR-026}, and to proposed detectors for FCC-hh~\cite{FCC:2018vvp} and the Muon Collider~\cite{InternationalMuonCollider:2024jyv}.

\begin{figure}[t]
    \begin{subfigure}[t]{0.45\textwidth}
        \centering
        \includegraphics[width=\textwidth]{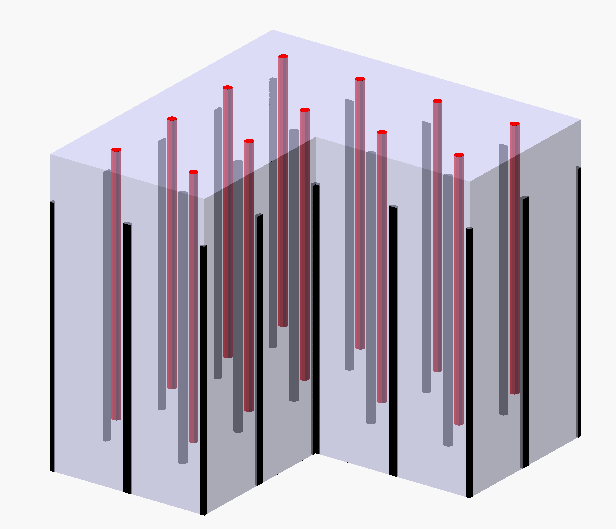}
        \caption{\label{fig:schematic-view}
            Diagrammatic illustration of a three-dimensional diamond sensor, depicting a segment comprising 
            four by four fundamental units. 
            The figure show in red electrodes connected to the polarization voltage, and in black the ones 
            grounded.From \cite{anderlini2025optimization3ddiamonddetectors}
        }
    \end{subfigure}
    \hfill
    \begin{subfigure}[t]{0.45\textwidth}
        \centering
        \includegraphics[width=\textwidth]{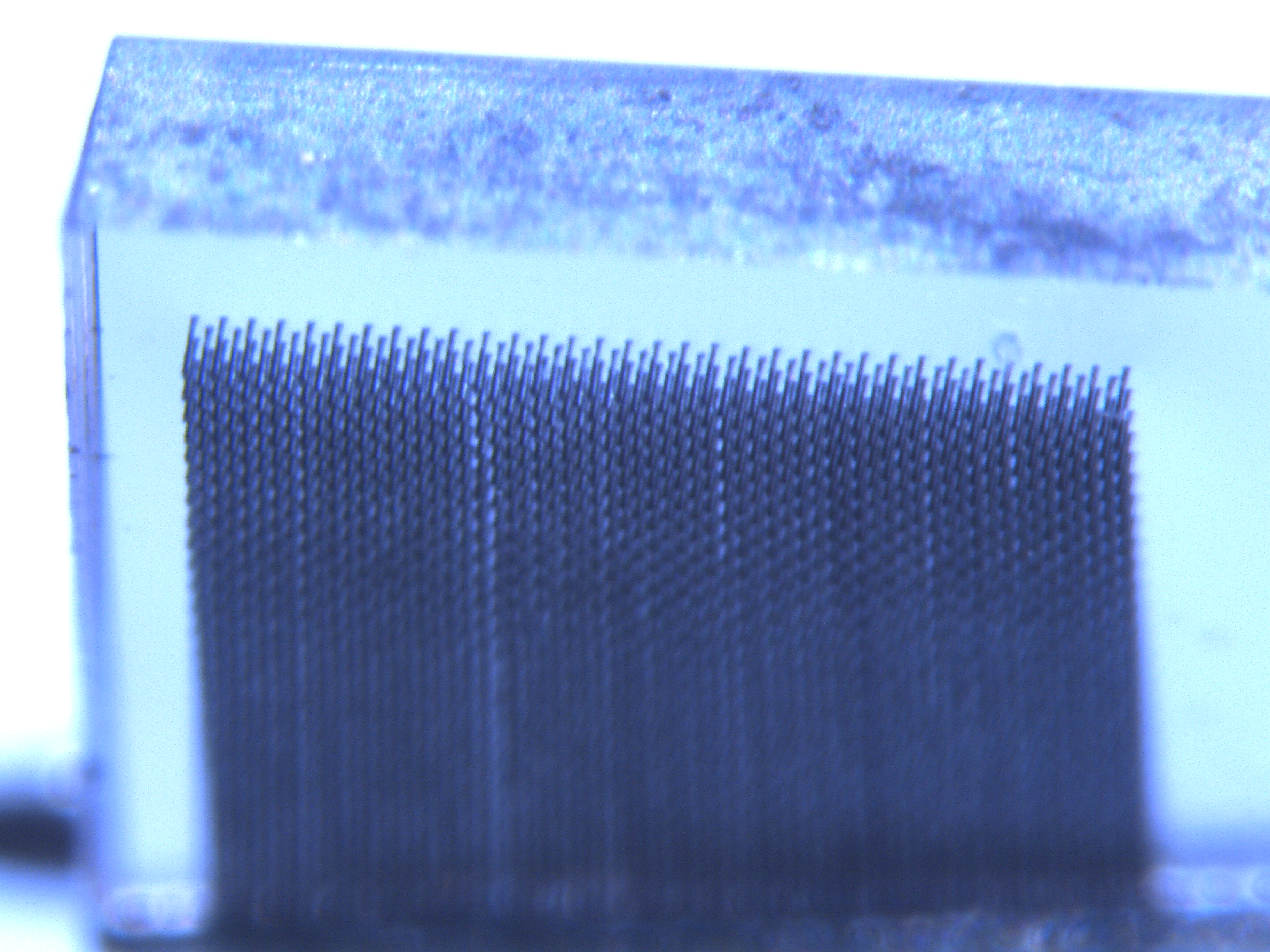}
        \caption{\label{fig:photo-sensor}
            A microscopic image of a 3D diamond sensor. 
            The specimen was tilted during acquisition to show the graphitised electrodes. From \cite{anderlini2025optimization3ddiamonddetectors}
        }
    \end{subfigure}
\end{figure}

Diamond detectors are promising candidates due to their superior radiation hardness and water-equivalence, making them ideal for dosimetry. Moreover, laser-induced graphitisation enables embedding conductive regions within the diamond bulk, allowing flexible electrode geometries~\cite{Oliva:2019alx,PORTER2023109692,Anderlini:2021pei,Bachmair:2015iba}. In \cref{fig:schematic-view} we report the schematic representation of the geometry of 3D diamond sensors, in which the graphite electrodes are orthogonal to the diamond surface. In \cref{fig:photo-sensor}, we present an actual picture of such diamond detector.


Simulating these devices is challenging: charge transport in the semiconductor and signal propagation through resistive electrodes contribute comparably to time resolution. A time-dependent extension of the Ramo-Shockley theorem captures these effects via dynamic weighting potentials, governed by Maxwell equations  in quasi-static approximation~\cite{Rigler:2e004jh, Janssens:2890572, Janssens:2022jos}.

By using the extended form of the Ramo-Shockley theorem for conductive media, the contribution of the resistive material to the signal formation is included in the time evolution of the weighting potential, which is the solution of the Maxwell equations in the quasi-static limit \cite{anderlini2025optimization3ddiamonddetectors}:
\begin{align}
    \epsilon \nabla^2 V(t, \mathbf{x}) &= - \rho (t, \mathbf{x}) \\
    \partial_t \rho (t, \mathbf{x}) &= \nabla \cdot \left[ \sigma(\mathbf{x}) \nabla V(t, \mathbf{x}) \right] \,,
\end{align}
where $V(t, \mathbf{x})$ is the weighting potential field, $\epsilon = \epsilon_0 \epsilon_r$ is the dielectric constant in the material, which is the same for diamond and graphite, and thus constant on the whole geometry, $\rho (t, \mathbf{x})$ is the charge distribution field, and $\sigma(\mathbf{x})$ is the function defining the conductivity in the diamond/graphite geometry. These equations can be condensed in a single, third-order partial differential equation
\begin{equation}\label{eq.PDE}
    \epsilon \partial_t \nabla^2 V(t, \mathbf{x}) +  \nabla \cdot \left[ \sigma(\mathbf{x}) \nabla V(t, \mathbf{x}) \right] = 0\,.
\end{equation}

The full system to describe the behaviour of the weighting field in the geometry is equipped with an initial condition (IC)
\begin{equation}
    V(t=0, \mathbf{x}) = \begin{cases}
        + V_0\,, &\mathbf{x} \in \partial C_+ \cap \partial \Omega \,,\\
        0\,, & \text{otherwise} \,,
    \end{cases}
\end{equation}
and boundary conditions (BC)
\begin{align}
    V(t, \mathbf{x}) &= +V_0\,, \quad \mathbf{x} \in \partial C_+ \cap \partial \Omega\,,\\
    V(t, \mathbf{x}) &= 0 \,,  \,\,\qquad  \mathbf{x} \in \partial C_- \cap \partial \Omega\,,\\
    \hat{\mathbf{n}} \cdot \nabla V(t, \mathbf{x}) &= 0\,, \qquad \,\, \mathbf{x} \in \partial\Omega \setminus\bigcup_iC_{i} \,,
\end{align}
where $\Omega$ is the whole diamond+graphite geometry, $C_\pm$ are the conductive cylinders attached to potential ($C_+$) or grounded ($C_-$). The last condition applies to all the boundary region of $\Omega$ where there is no intersection with conductive columns.  

Notice that this equation accounts for the timely response of the detector, with respect to its design: the manufacturable aspect of the geometry is kept into account in the function $\sigma(\mathbf{x})$, which is the result of the manufacturing process. 

The computation needed to solve the differential equation describing the system to obtain the time-dependent electric field, via the (reformulated) Ramo-Shockley theorem, are highly non-trivial. Using standard numerical solver, like COMSOL finite element method~\cite{comsol} on a custom mesh~\cite{Janssens:2890572}, or spectral methods \cite{anderlini2025optimization3ddiamonddetectors}, gives mesh-ful results, on a grid-spaced time steps.

This  motivates the role that can be played by Physics Informed Neural Networks (PINNs) \cite{raissi2017physics, raissi2019physics}, either as a surrogate model as well as a mixed numerical solver. In fact, we may leverage the power of neural networks to extrapolate statistical relations among data, and, by adding physical information about the system through the loss as prescribed by PINN, create a parametric surrogate model to optimize the design of Diamond detectors.

\section{Methods: using physics informed neural networks to solve the governing PDE in a \textit{meshless} way}
\textit{Physics-Informed Neural Networks} are neural networks (NNs) whose training is based on the mathematical model that governs the physical phenomena that we are studying, and, possibly, on available data. More precisely, in our case, rewriting the PDE \eqref{eq.PDE} with BC and IC in a more compact form as\footnote{With the notation $\partial_\bullet$ we denote any derivative with respect to either time or space.}
\begin{equation*}
\left\{
\begin{array}{ll}
\mathcal{F}(V,\partial_\bullet V, \partial_\bullet^2 V; t, \mathbf{x})=0,&\quad (t,\mathbf{x})\in \text{$\mathbb{R} \times \Omega$} \\
\mathcal{B}(V,\partial_\bullet V; t, \mathbf{x})=0,&\quad (t,\mathbf{x})\in \text{ $\mathbb{R} \times \partial\Omega$},\\
\mathcal{I}(V; t, \mathbf{x})=0,&\quad (t,\mathbf{x})\in  \text{$\{0\} \times \overline{\Omega}$},
\end{array}\right.
\end{equation*}
we optimize the parameters of the NN by minimizing the following loss function:
\begin{align*}
\mathcal{L}_{\text{tot}}&=\lambda_{\text{data}} \mathcal{L}_{\text{data}}\big(V_\theta(t,\mathbf{x}),V_{\text{data}}\big)+\lambda_{\text{PDE}} \mathcal{L}_{\text{PDE}}\big(V_\theta(t,\mathbf{x})\big)  + \lambda_{\text{BC}} \mathcal{L}_{\text{BC}}\big(V_\theta(t,\mathbf{x})\big)+\lambda_{\text{IC}} \mathcal{L}_{\text{IC}}\big(V_\theta(t,\mathbf{x})\big),
\end{align*}
where
\begin{align*}
\mathcal{L}_{\text{data}}&=\frac{1}{N_d}\,\sum_{i=1}^{N_d}\left|V_\theta(t^i_d,\mathbf{x}^i_d)-V_{\text{data}}^i\right|^2,\quad
\mathcal{L}_{\text{PDE}}=\frac{1}{N_P}\,\sum_{i=1}^{N_P}\left|\mathcal{F}(V_\theta,\partial_\bullet V_\theta, \partial_\bullet^2 V_\theta)(t_P^i,\mathbf{x}^i_P)\right|^2,\\
\mathcal{L}_{\text{BC}}&=\frac{1}{N_B}\,\sum_{i=1}^{N_B}\left|\mathcal{B}(V_\theta,\partial_\bullet V_\theta)(t_B^i,\mathbf{x}^i_B)\right|^2,\quad
\mathcal{L}_{\text{IC}}=\frac{1}{N_I}\,\sum_{i=1}^{N_I}\left|\mathcal{I}(V_\theta)(t_I^i,\mathbf{x}^i_I)\right|^2.
\end{align*}
Here, the set $\left\{t_d^i,\mathbf{x}^i_d,V_{\text{data}}^i\right\}_{i=1,\ldots,N_d}$ denotes the available data, and $\left\{t^i_P,\mathbf{x}_P^i\right\}_{i=1,\ldots,N_P} $ $\in \mathbb{R}\times \Omega$, $\left\{t^i_B,\mathbf{x}_B^i\right\}_{i=1,\ldots,N_B} $ $\in \mathbb{R}\times \partial \Omega$, $\left\{t_I^i,\mathbf{x}_I^i\right\}_{i=1,\ldots,N_I} \in \{0\}\times \overline{\Omega}$ are random-generated points for PDE, BC, IC validation of NN, respectively.
Moreover, $\lambda_{\text{data}}, \lambda_{\text{PDE}}, \lambda_{\text{BC}}, \lambda_{\text{IC}}$ are the weights of the respective loss functions. Finally, $\theta$ denotes the NN trainable weights.

\begin{figure}
    \centering
    \includegraphics[width=0.75\linewidth]{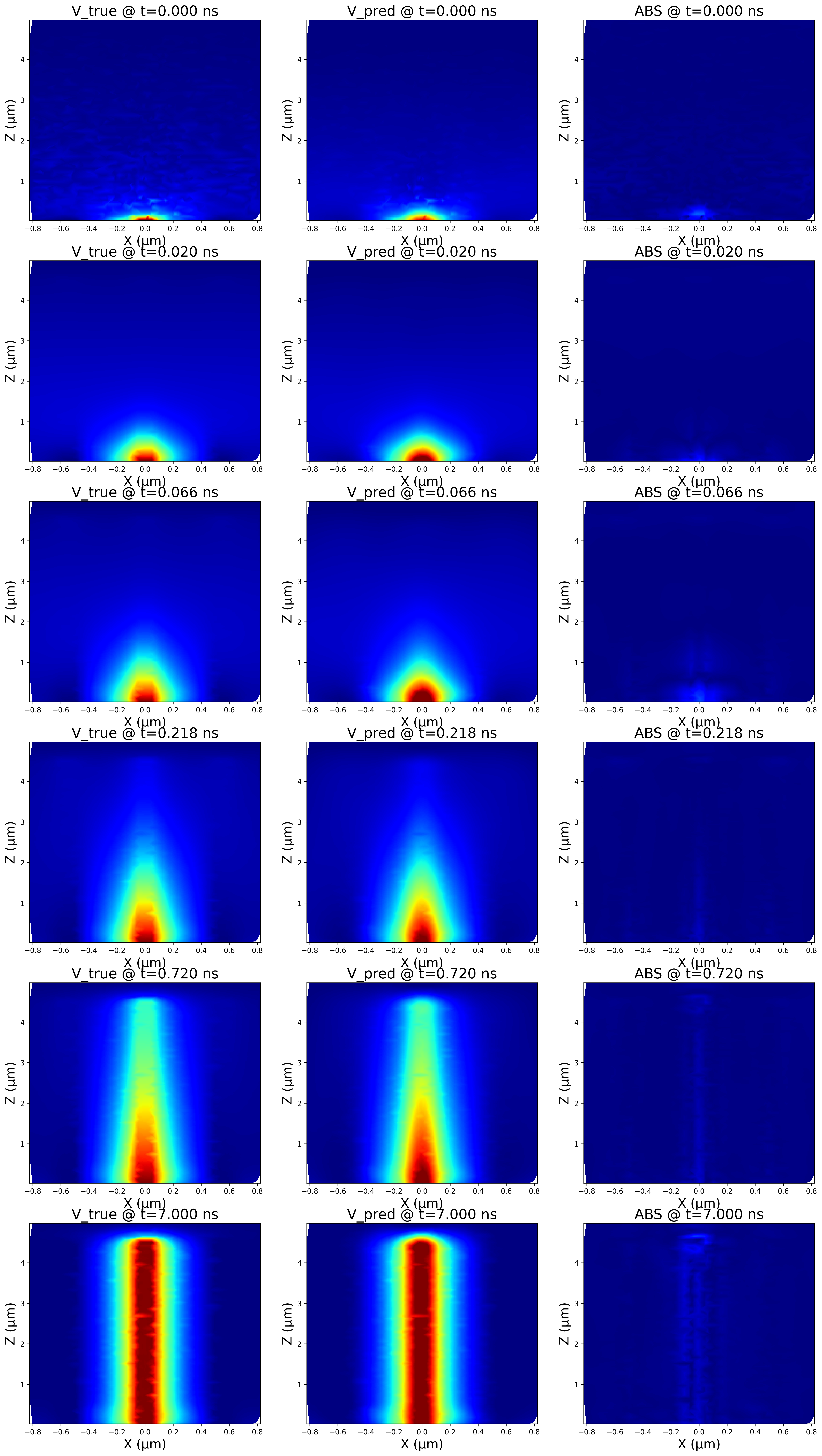}
    \caption{Results of the trained model. The first row represents the real data, i.e. the data obtained from the FEM simulation. The second row represents the MoE network prediction. The last row is the absolute error. }
    \label{fig:valplot}
\end{figure}

To tackle this problem, we designed a Mixture-of-Experts (MoE) model \cite{MoePinn}, comprising three experts: two are Multi-Layer Perceptron (MLP) architectures, with 6 layers and 256 nodes per layer, with skip connections and adaptive activations \cite{jagtap2020adaptive}, one uses sigmoid-weighted linear unit (\texttt{SiLU}) \cite{elfwing2018sigmoid} as activation function, while the other uses the self-scalable tanh (\texttt{STAN}) \cite{gnanasambandam2022self}; the third architecture is a Multi-scale Fourier Network \cite{WANG2021113938}, again with 6 layers and 256 nodes per layer, skip connections, adaptive activations and \texttt{SiLU} activation function. The gate network is a simply 2 layer, 128 node per layer network again with 6 layers and 256 nodes per layer, skip connections, adaptive activations and \texttt{SiLU} activation function. The role of the gate network is to produce, for each input data $(t_i, \mathbf{x}_i)$ an \textit{importance vector} $\mathbf{g} : g_i, i=1,\ldots,N_{\text{experts}}$ to weight the contribution of each expert $u_i$, so that the whole solution is 
\begin{equation}
    V_\theta (t_i, \mathbf{x}_i) = \sum_{i=1}^{N_{\text{experts}}} g_i (t_i, \mathbf{x}_i) \,  u_i  (t_i, \mathbf{x}_i) \,.
\end{equation}


To draw the PDE, BC and IC data we have used a \textit{quasi-random} sampling using Halton sequences \cite{halton1960efficiency}. Furthermore, we have used an importance measure \cite{nabian2021efficient} for the field, where the importance function $i(t, \mathbf{x})$ was
\begin{equation}
    i(t, \mathbf{x}) = \sqrt{ \left(\nabla V_\theta (t, \mathbf{x})\right)^2 + 5 \left(\partial_t V_\theta (t, \mathbf{x})\right)^2 } +  \left| V_\theta (t, \mathbf{x})\right| + 10 \,,
\end{equation}
so that we can have a focus on region with high field values and/or fast changing values. Finally, we have used also a temporal loss weighting for the PDE, 
\begin{equation}
    \lambda_{\text{PDE}} (t) = 1 + c_T \left( 1 - \frac{t}{T} \right) \,,
\end{equation}
to focus on the initial stages of the dynamics. Finally, to adaptively weight the multi-objective loss, we used the Neural Tangent Kernel approach \cite{wang2022and}.

\begin{figure}[ht]
    \centering
    \includegraphics[width=0.55\linewidth]{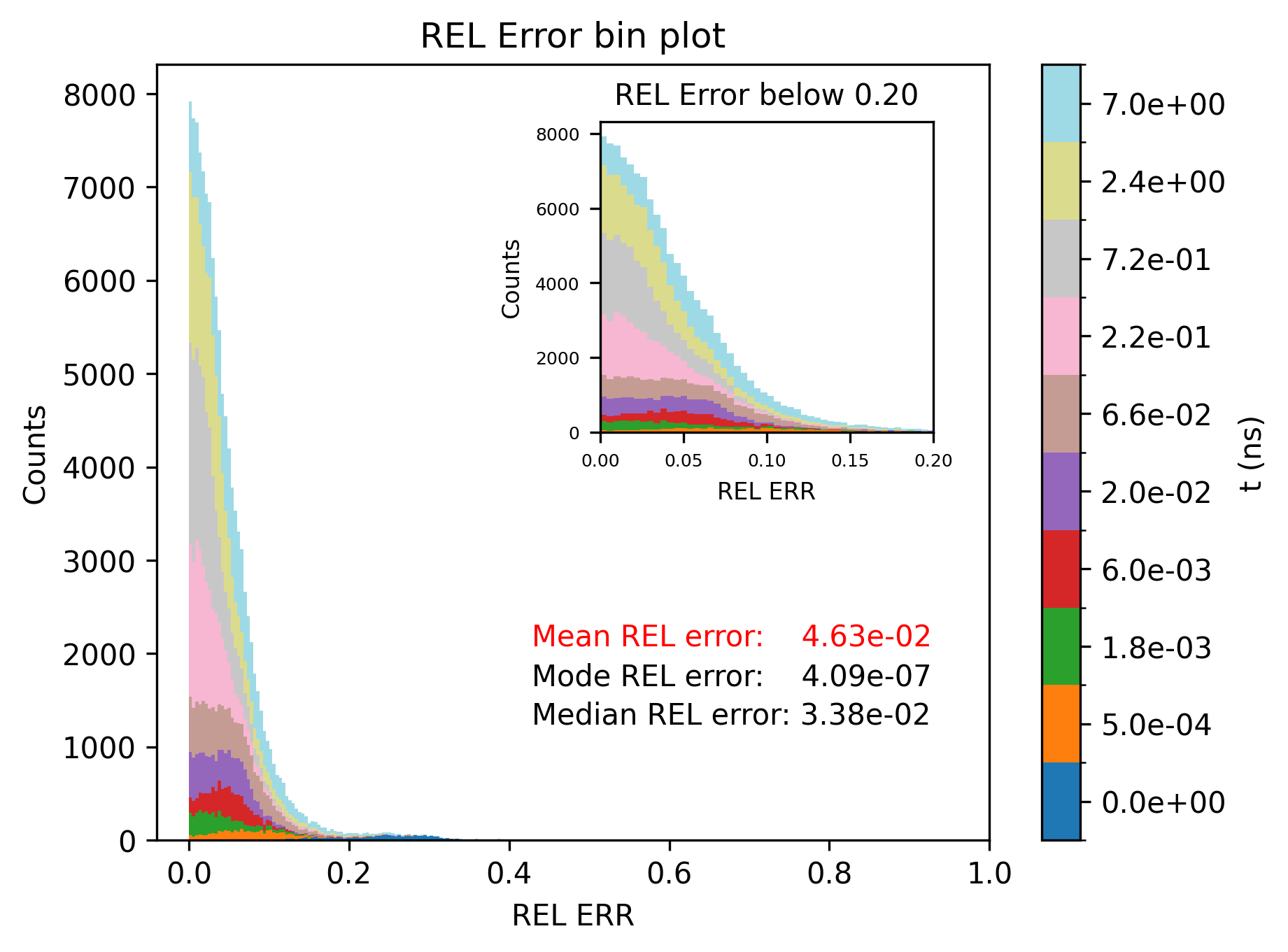}
    \caption{Histogram reporting the Relative Errors, i.e.~the L1 distance over the true value of the weighting field,  between the MoE-PINN prediction and the data points. In the inlet, it is reported the zoom of the histogram plot with error below 20\%.}
    \label{fig:hist-res}
\end{figure}

\section{Results}

The implementation of the code needed to perform this approach has been done using the Python language, the PyTorch-based open-source package Nvidia PhysicsNemo\footnote{\url{https://github.com/NVIDIA/physicsnemo-sym}.} (formerly Modulus), and the development, debugging, training and test process has been conducted on the HPC cloud-based platform offered by AI\_INFN \cite{petrini2025developing}. 

The results obtained are reported in \cref{fig:valplot} and \cref{fig:hist-res}. In \cref{fig:valplot} we report the time evolution of $V$ in a few time steps (not seen during training) at $y=0 \ \mu m$. The first column represent the "true" field (i.e., the FEM simulation data); the second column represent the MoE-PINN prediction; the last column, the absolute error. From here we see that the MoE-PINN is capable of inferring the dynamics of the system.

\begin{table}[h]
\centering
\small
\begin{tabular}{|
>{\columncolor[HTML]{ECF4FF}}c |
>{\columncolor[HTML]{FFFFFF}}c 
>{\columncolor[HTML]{FFFFFF}}c 
>{\columncolor[HTML]{FFFFFF}}c |}
\hline
\cellcolor[HTML]{FFFFFF}Relative Error / Time & \cellcolor[HTML]{FFFFC7}Mean & \cellcolor[HTML]{FFFFC7}Median & \cellcolor[HTML]{FFFFC7}Mode (bin n$^{\circ}$) \\ \hline
$5.0\cdot 10^{-4}$ ns                         & $8.4\,\%$                    & $7.5\,\%$                      & $6.4\,\% \ \ (16)$                                \\
$1.8\cdot 10^{-3}$ ns                         & $4.2 \,\%$                   & $3.4 \,\%$                     & $1.2\,\% \ \ (3)$                                 \\
$6.0\cdot 10^{-3}$ ns                         & $4.9\,\%$                    & $4.8\,\%$                      & $3.6\,\% \ \ (9)$                                 \\
$2.0\cdot 10^{-2}$ ns                         & $4.4 \,\%$                   & $4.2 \,\%$                     & $3.6\,\% \ \ (9)$                                 \\
$6.6\cdot 10^{-2}$ ns                         & $5.4 \,\%$                   & $4.7\,\%$                      & $4.1\cdot 10^{-5}\,\% \ \ (0)$                    \\
$2.2\cdot 10^{-1}$ ns                         & $3.0 \,\%$                   & $2.3\,\%$                      & $0.1\% \ \ (2)$                                   \\
$7.2\cdot 10^{-1}$ ns                         & $2.9 \,\%$                   & $2.2 \,\%$                     & $4.1\cdot 10^{-5}\,\% \ \ (0)$                    \\
$2.4$ ns                                      & $3.6 \,\%$                   & $2.9 \,\%$                     & $4.1\cdot 10^{-5}\,\% \ \ (0)$                    \\
$7.0$ ns                                      & $6.0 \,\%$                   & $5.2 \,\%$                     & $4.1\cdot 10^{-5}\,\% \ \ (0)$                    \\ \hline
\end{tabular}
\caption{List of errors per integration time step. The mode is computed as the central value of the bin with most counts, in case of the histogram computed with 200 bins. }
\end{table}

In \cref{fig:hist-res} is reported the histogram plot of the Relative errors, separated by integration time step. We see that the relative error \textit{decrase} in time, and that the overall mean error is below $5\%$, the median error is around $3\%$, while the mode error is $0.00005\%$, i.e. the most counts overall are in the first bin (in a 200 bins histogram). 

\section{Conclusion} \label{sec:conclusion}
This work demonstrates the application of Physics-Informed Neural Networks (PINNs) to the design and optimization of diamond detectors. Specifically, we employed a Mixture-of-Experts PINN (MoE-PINN) to perform mesh-free interpolation of Maxwell’s equations, under quasi-static approximation, both spatially and temporally, capturing the detector’s response to charged particle passage via an extended Ramo-Shockley theorem for resistive media.


Despite the complexity of the governing third-order PDE, characterised by a third-order PDE with an underlying approximate spatial functional symmetry\footnote{Which is valid everywhere except where $\nabla \sigma \cdot \nabla V \neq 0$, i.e., only on the conductive cylinders surface.} $V(t, \mathbf{x})  \mapsto V(t, \mathbf{x}) + f(\mathbf{x})$ if $ \nabla^2 f(\mathbf{x}) = 0$, we successfully trained the MoE-PINN as a physics-informed, mesh-free surrogate for the numerical solver. The model achieved a median error of approximately 3\%, with the most frequent error mode being lower at almost each timesteps.

\section*{Acknowledgements}


\paragraph{Funding information}
This work is partly supported by ICSC – Centro Nazionale di Ricerca in High Performance Computing, Big Data and Quantum Computing, funded by European Union – NextGenerationEU.

The work of {AB} and AR was funded by Progetto ICSC - Spoke 2 - Codice CN00000013 - CUP I53C21000340006 - Missione 4 Istruzione e ricerca - Componente 2 Dalla ricerca all'impresa – Investimento 1.4.

\bibliography{bibliography.bib}










\end{document}